\documentclass{PoS}
\usepackage{amsmath} 
\usepackage{pstricks}
\usepackage{longtable}
\usepackage{color}
\usepackage{amssymb}
\usepackage{cite}

\input{epsf} 

\newcommand\as{\alpha_{\mathrm{S}}} 
\newcommand\f[2]{\frac{#1}{#2}} 
\def\beq{\begin{equation}} 
\def\eeq{\end{equation}} 
\def\to{\rightarrow} 
 
\def\beeq{\begin{eqnarray}}
\def\eeeq{\end{eqnarray}}

\def\mur{\mu_R} 
\def\muf{\mu_F}
\def\mur2{\mu_R^2} 
\def\muf2{\mu_F^2}

\title{
\vspace*{-3.7cm}
\begin{minipage}{\textwidth}
{\normalfont\small DESY 14-080 / LPN 14-073
\hspace{\fill} May 2014
}\\
\end{minipage}\\[60pt]
  Next-to-Next-to-Leading Order QCD Corrections to Higgs Boson Pair Production}

\ShortTitle{NNLO QCD Corrections to Higgs Boson Pair Production}

\author{Daniel de Florian$^a$ and \speaker{Javier Mazzitelli}$^{a,b}$%
         \\
        \llap{$^a$}Departamento de F\'\i sica, FCEyN, Universidad de Buenos Aires \\
   (1428) Pabell\'on 1, Ciudad Universitaria, Capital Federal, Argentina\\
   \llap{$^b$}Deutsches Elektronensynchrotron DESY \\
   Platanenallee 6, D--15738 Zeuthen, Germany\\
        E-mail: \email{deflo@df.uba.ar},\email{jmazzi@df.uba.ar}}

\abstract{
We present the Higgs boson pair production cross section  at next-to-next-to-leading order in QCD within the large top-mass approximation.
Numerical results for the LHC are provided, finding an increase of ${\cal O}(20\%)$ with respect to the previous order prediction and a substantial reduction in the scale dependence.
We normalize our results using the full top- and bottom-mass dependence at leading order.
}

\FullConference{ Loops and Legs in Quantum Field Theory - LL 2014,\\
		 27 April - 2 May 2014 \\
		 Weimar, Germany }

\begin{document}

\section{INTRODUCTION}

Recently, both ATLAS \cite{Aad:2012tfa} and CMS \cite{Chatrchyan:2012ufa} collaborations discovered a new boson at the Large Hadron Collider (LHC), whose properties are so far compatible with the long sought standard model (SM) Higgs boson \cite{Englert:1964et,Higgs:1964pj,Higgs:1964ia}.
In order to decide whether this particle is indeed responsible for the electroweak symmetry breaking, a precise measurement of its couplings to fermions, gauge bosons and its self-interactions is needed.
In particular, the knowledge of the Higgs self-couplings is the only way to reconstruct the scalar potential.

The possibility of observing Higgs pair production at the LHC have been discussed in Refs. \cite{Baur:2002qd,
Dolan:2012rv,Papaefstathiou:2012qe,
Baglio:2012np,Baur:2003gp,Dolan:2012ac,Goertz:2013kp,
Shao:2013bz,Gouzevitch:2013qca}.
In general, it has been shown that despite the smallness of the signal and the large background its measurement can be achieved at a luminosity-upgraded LHC.

The dominant mechanism for SM Higgs pair production at hadron colliders is gluon fusion, mediated by a heavy-quark loop.
The leading-order (LO) cross section has been calculated in Refs. \cite{Glover:1987nx,Eboli:1987dy,Plehn:1996wb}.
The next-to-leading order (NLO) QCD corrections have been evaluated in Ref. \cite{Dawson:1998py} within the large top-mass approximation and found to be large, with an inclusive $K$ factor close to $2$.
The finite top-mass effects were analysed at this order in Ref. \cite{Grigo:2013rya}, finding that a precision of ${\cal O}(10\%)$ can be achieved if the exact top-mass leading-order cross section is used to normalize the corrections.

Given the size of the NLO corrections, it is necessary to reach higher orders to provide accurate theoretical predictions.
In this proceeding we present the next-to-next-to-leading order (NNLO) corrections for the inclusive Higgs boson pair production cross section\cite{deFlorian:2013jea}.

\section{DESCRIPTION OF THE CALCULATION}
The effective single and double-Higgs coupling to gluons is given, within the large top-mass approximation, by the following Lagrangian
\beq
{\cal L}_{\text{eff}}=
-\f{1}{4}G_{\mu\nu}G^{\mu\nu}
\left(C_H\f{H}{v}-C_{HH}\f{H^2}{v^2}\right)\,.
\eeq
Here $G_{\mu\nu}$ stands for the gluonic field strength tensor and $v\simeq 246\,\text{GeV}$ is the Higgs vacuum expectation value.
While the ${\cal O}(\as^3)$ of the $C_{H}$ expansion is known \cite{Kramer:1996iq,Chetyrkin:1997iv}, the QCD corrections of $C_{HH}$ are only known up to ${\cal O}(\as^2)$ \cite{Djouadi:1991tka}. Up to that order, both expansions yield the same result.

The NNLO contributions to the SM Higgs boson pair production squared matrix element can be separated into two different classes: (a) those containing two gluon-gluon-Higgs vertices (either $ggH$ or $ggHH$) and (b) those containing three or four effective vertices.
Given the similarity between $ggH$ and $ggHH$ vertices, the contributions to the class (a) are equal to those of single Higgs production, except for an overall LO normalization (assuming that $C_H=C_{HH}$ up to ${\cal O}\left(\as^3\right)$). These results can be obtained from Refs. \cite{Harlander:2002wh,Anastasiou:2002yz,Ravindran:2003um}.

\begin{figure}
\begin{center}
\begin{tabular}{c c}
\epsfxsize=6.8truecm
\epsffile{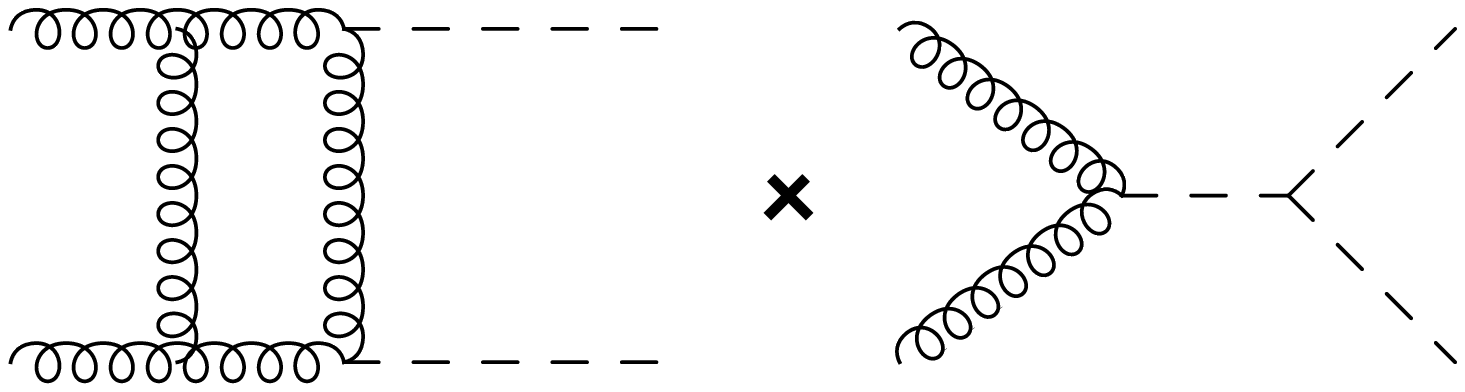}
\hspace{0.5cm}
&
\epsfxsize=6.8truecm
\epsffile{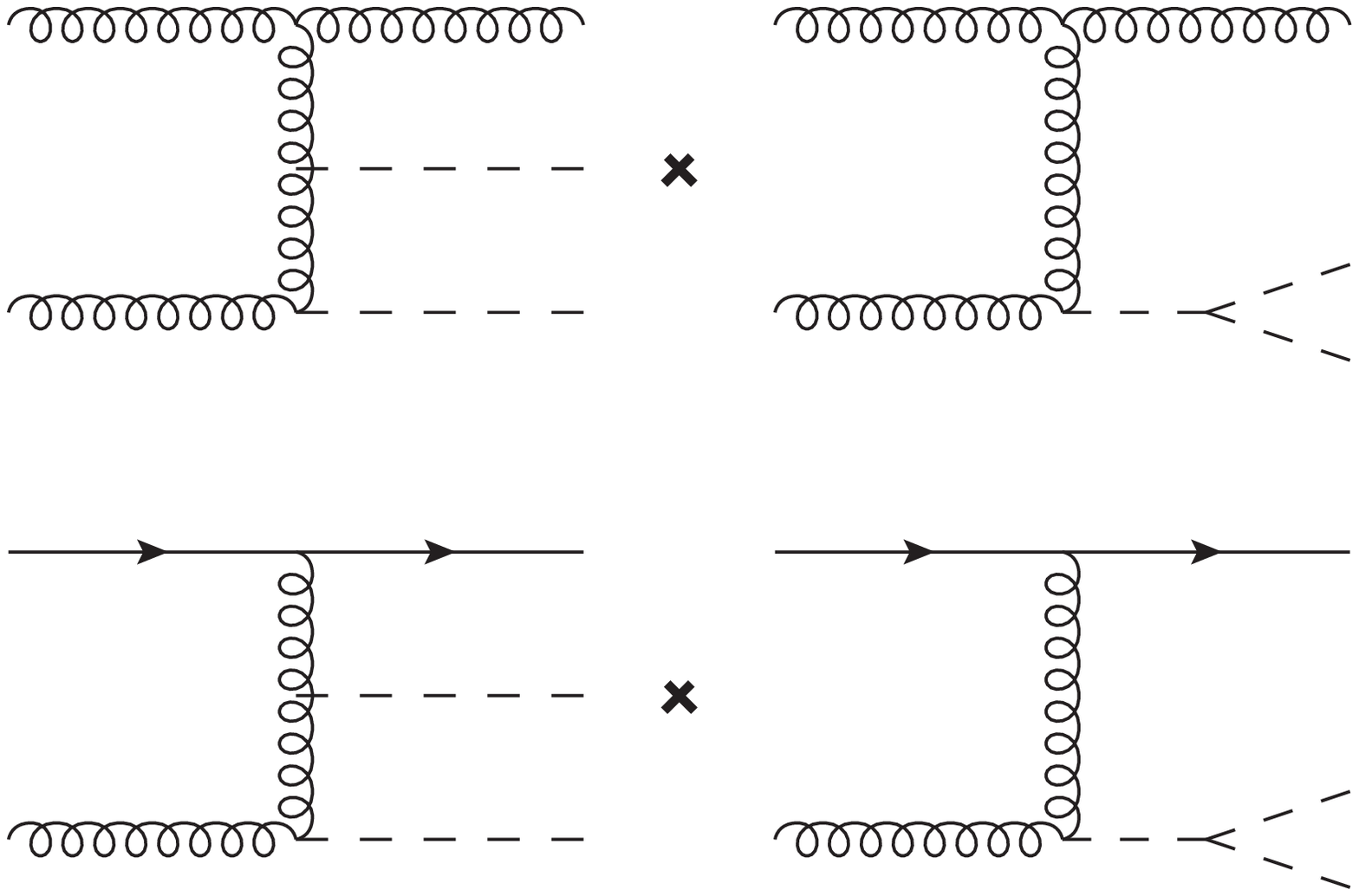}\\
\end{tabular}
\end{center}
\vspace{-0.5cm}
\caption{\label{diagramas}
Example of Feynman diagrams needed for the NNLO calculation for the virtual corrections (left) and the real corrections (right) for $gg\to HHg$ (top) and $qg\to HHq$ (bottom) subprocesses.  Other parton subprocesses can be obtained from crossings.}
\end{figure}

Contributions to the class (b) first appear at NLO as a tree-level contribution to the subprocess $gg\to HH$, given that each $ggH$ and $ggHH$ vertex is proportional to $\as$.
Then, at NNLO we have one-loop corrections and single real emission corrections.
The virtual corrections only involve the gluon initiated partonic channel.
The remaining contributions involve the partonic subprocesses $gg\to HH+g$ and $qg\to HH+q$ (with the corresponding crossings). Examples of the Feynman diagrams involved in the calculation are shown in Figure \ref{diagramas}.

For both virtual and real corrections, we used the {\sc Mathematica} packages {\sc FeynArts} \cite{Hahn:2000kx} and {\sc FeynCalc} \cite{Mertig:1990an} in order to generate the Feynman diagrams and evaluate the corresponding amplitudes.
The calculation was performed using nonphysical polarizations, which we cancel by including ghosts in the initial and final states.
We used the {\sc Fire} algorithm \cite{Smirnov:2008iw} to reduce the virtual contributions into master integrals, which were obtained from Ref. \cite{Ellis:2007qk}.
For the real emission processes we used the Frixione, Kunszt, and Signer subtraction method \cite{Frixione:1995ms} in order to subtract the soft and collinear divergencies. Further details of the calculation, together with the explicit expressions for the NNLO results, can be found in Refs. \cite{deFlorian:2013uza,deFlorian:2013jea}.

\section{PHENOMENOLOGY}

Here we present the numerical results for the LHC. At each order, we use the corresponding MSTW2008 \cite{Martin:2009iq} set of parton distributions and QCD coupling.
We recall that we always normalize our results using the exact top- and bottom-mass dependence at LO.
For this analysis we use $M_H=126\,\text{GeV}$, $M_t=173.18\,\text{GeV}$ and $M_b=4.75\,\text{GeV}$.
The bands of all the plots are obtained by varying independently the factorization and renormalization scales in the range $0.5\,Q\leq \mu_F,\mu_R \leq 2\,Q$, with the constraint $0.5\leq \mu_F/\mu_R \leq 2$, being $Q$ the invariant mass of the Higgs pair system.

We assume for the phenomenological results that the two-loop corrections to the effective vertex $ggHH$  are the same than those of $ggH$ (that is $C_{HH}^{(2)}=C_{H}^{(2)}$, following the notation of Ref. \cite{deFlorian:2013uza}), as it happens at one-loop order.
We change  its value in the range $0\leq C_{HH}^{(2)}\leq 2C_{H}^{(2)}$ in order to evaluate the impact of this unknown coefficient and find a variation in the total cross section of less than $2.5\%$.

\begin{figure}
\begin{center}
\begin{tabular}{c}
\epsfxsize=8.7truecm
\epsffile{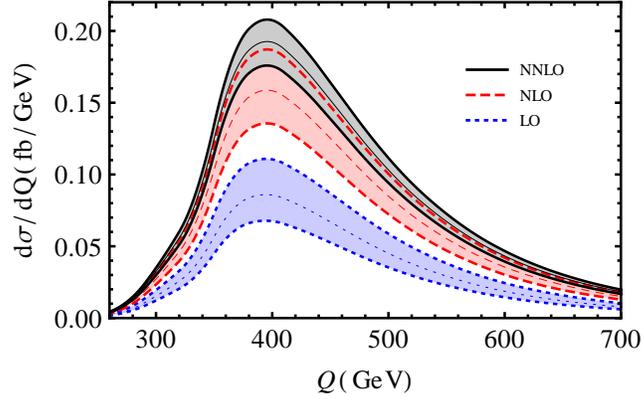}\\
\end{tabular}
\end{center}
\vspace{-0.7cm}
\caption{\label{q2distr}
Higgs pair invariant mass distribution at LO (dotted blue), NLO (dashed red) and NNLO (solid black) for the LHC at c.m. energy $E_{cm}=14\,\text{TeV}$. The bands are obtained by varying $\mu_F$ and $\mu_R$ in the range $0.5\,Q\leq \mu_F,\mu_R \leq 2\,Q$ with the constraint $0.5\leq \mu_F/\mu_R \leq 2$.}
\end{figure}

In Figure \ref{q2distr} we present the LO, NLO and NNLO predictions for the hadronic cross section at the LHC as a function of the Higgs pair invariant mass, for a c.m. energy $E_{cm}=14\,\text{TeV}$. As can be noticed from the plot, only at this order the first sign of convergence of the perturbative series appears, finding a nonzero overlap between the NLO and NNLO bands.
Second order corrections are sizeable, this is noticeable already at the level of the total inclusive cross sections,
where the increase with respect to the NLO result is of ${\cal O}(20\%)$, and the $K$ factor with respect to the LO prediction is about $K_{\text{NNLO}}=2.3$.
The scale dependence is clearly reduced at this order, resulting in a variation of about $\pm8\%$ around the central value, compared to a total variation of ${\cal O}(\pm20\%)$ at NLO.

\begin{figure}
\begin{center}
\begin{tabular}{c}
\epsfxsize=8.7truecm
\epsffile{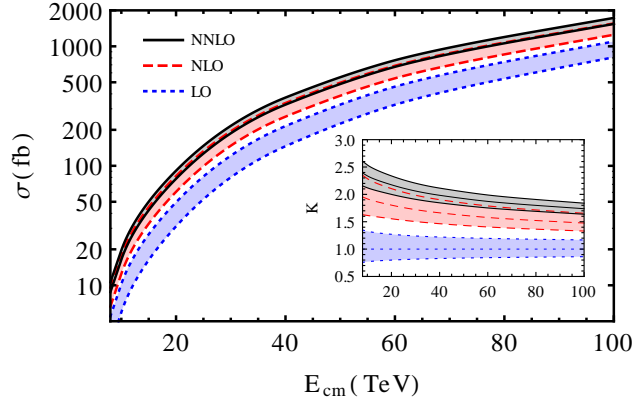}\\
\end{tabular}
\end{center}
\vspace{-0.7cm}
\caption{\label{sh}
Total cross section as a function of the c.m. energy $E_{cm}$ for the LO (dotted blue), NLO (dashed red) and NNLO (solid black) prediction. The bands are obtained by varying $\mu_F$ and $\mu_R$ as indicated in the main text. The inset plot shows the corresponding $K$ factors.}
\end{figure}

In Figure \ref{sh} we show the total cross section as a function of the c.m. energy $E_{cm}$, in the range from $8\,\text{TeV}$ to $100\,\text{TeV}$. We can observe that the size of the NLO and NNLO corrections is smaller as the c.m. energy increases. We can also notice that the scale dependence is substantially reduced in the whole range of energies when we include the second order corrections.
The ratio between NNLO and NLO predictions as a function of the c.m. energy is quite flat, running from $1.22$ at $8\,\text{TeV}$ to $1.18$ at $100\,\text{TeV}$.
On the other hand, the ratio between NNLO and LO runs from $2.39$ to $1.74$ in the same range of energies.

Finally, we present in Table \ref{tabla} the value of the NNLO cross section for $E_{cm}=8$, 14, 33 and $100\,\text{TeV}$.
We have taken into account three sources of theoretical uncertainties: missing higher orders in the QCD perturbative expansion, which are estimated by the scale variation, and uncertainties in the determination of the parton flux and strong coupling.
To estimate the parton dinstributions and coupling constant uncertainties we used the MSTW2008 $90\%$ C.L. error PDF sets \cite{Martin:2009bu}, which are known to provide very close results to the PDF4LHC working group recommendation for the envelope prescription \cite{Botje:2011sn}. As we can observe from Table \ref{tabla}, nonperturbative and perturbative uncertainties are of the same order.

\begin{table}
\begin{center}
\begin{tabular}{l  c  c  c  c }
\hline\hline
$E_{cm}$ & $8\text{ TeV}$ & $14\text{ TeV}$ & $33\text{ TeV}$ & $100\text{ TeV}$ \\
\hline
$\sigma_{\text{NNLO}}$ & $9.76\text{ fb}$ & $40.2\text{ fb}$ & $243\text{ fb}$ & $1638\text{ fb}$ \\
Scale $[\%]$ &\footnotesize $+9.0-9.8\,$ &\footnotesize $\,+8.0-8.7\,$ &\footnotesize $\,+7.0-7.4\,$ &\footnotesize $\,+5.9-5.8$ \\ 
PDF $[\%]$ &\footnotesize $+6.0-6.1\,$ &\footnotesize $\,+4.0-4.0\,$ &\footnotesize $\,+2.5-2.6\,$ &\footnotesize $\,+2.3-2.6$ \\ 
PDF$+\as$  $[\%]$ &\footnotesize $+9.3-8.8\,$ &\footnotesize $\,+7.2-7.1\,$ &\footnotesize $\,+6.0-6.0\,$ &\footnotesize $\,+5.8-6.0$ \\ 
\hline\hline
\end{tabular}
\caption{Total cross section as a function of the c.m. energy at NNLO accuracy. We use the exact LO prediction to normalize our results. The different sources of theoretical uncertainties are discussed in the main text.\label{tabla}}
\end{center}
\end{table}

It is worth noticing that the soft-virtual approximation, which was presented in Ref. \cite{deFlorian:2013uza}, gives an extremely accurate prediction for the NNLO cross section, overestimating for example the $E_{cm}=14\,\text{TeV}$ result by less than $2\%$.
As expected, this approximation works even better than for single Higgs production, due to the larger invariant mass of the final state.

As was mentioned before, the finite top-mass effects were analysed in Ref. \cite{Grigo:2013rya}. They found that these terms provide an increase of ${\cal O}(10\%)$ in the NLO prediction at $E_{cm}=14\text{ TeV}$. This means that the NLO contribution separately is underestimated by about a $20\%$ in the large top-mass limit.
If this is the case also at the next order, then the finite top-mass contributions are expected to have an effect of  ${\cal O}(5\%)$ in the NNLO prediction (provided that the exact LO is used to normalize the results, and that the finite top-mass effects are included at NLO).
At higher collider energies the approximation is expected to be less accurate, given that larger invariant masses of the Higgs pair system are involved.

\begin{figure}
\begin{center}
\begin{tabular}{c}
\epsfxsize=8.7truecm
\epsffile{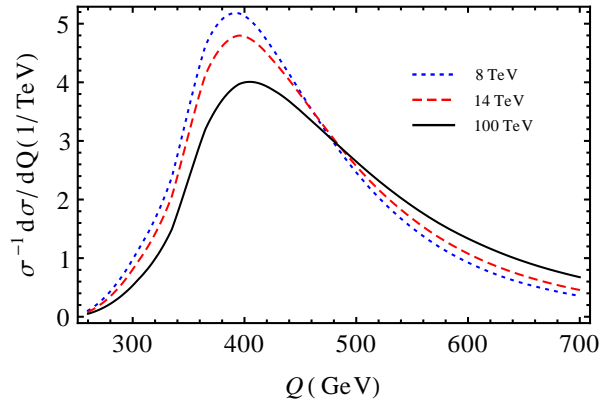}\\
\end{tabular}
\end{center}
\vspace{-0.7cm}
\caption{\label{Q2_sh}
Higgs pair invariant mass distribution at NNLO for the LHC at c.m. energy of $8\text{ TeV}$ (dotted blue), $14\text{ TeV}$ (dashed red) and $100\text{ TeV}$ (solid black), normalized by the total cross section.
}
\end{figure}

To analyse this statement, we show in Figure \ref{Q2_sh} the invariant mass distribution at NNLO for $E_{cm}=8$, $14$ and $100\text{ TeV}$, normalized by the total cross section.
We can observe that the maximum of the distribution is approximately in the same position, but the bulk of the the cross section shifts towards larger values of $Q$.
Nevertheless, this shift is not really dramatic: at $8$, $14$ and $100\text{ TeV}$ the $70\%$ of the inclusive cross section comes from $Q<485$, $505$ and $550\text{ GeV}$ respectively.
Then, the effective theory could be still reliable to compute the NNLO corrections for large collider energies.

\section*{ACKNOWLEDGEMENTS}
This work was supported in part by UBACYT, CONICET, ANPCyT and the Research Executive Agency (REA) of the European Union under the Grant Agreement number PITN-GA-2010-264564 (LHCPhenoNet).

\bibliography{PoS_LL14}

\end{document}